\documentclass[prl,twocolumn,showpacs,preprintnumbers,superscriptaddress,amsthm,amsmath,amssymb]{revtex4}
\usepackage{graphicx}
\usepackage{dcolumn}
\usepackage{bm}
\usepackage{color}
\usepackage{booktabs, longtable}
\usepackage{multirow}
\usepackage{ltxtable}
\usepackage{makecell}
\usepackage{float}
\usepackage{rotating}
\usepackage{dcolumn,siunitx,mathptmx,booktabs}
\usepackage{array}
\usepackage[colorlinks,linkcolor=blue,anchorcolor=blue,citecolor=red]{hyperref}

\begin{document}
\renewcommand\arraystretch{1.2}
\title{
Learning the structure of
giant resonances from their $\gamma$-decay
}
\author{W. L. Lv}
\affiliation{School of Nuclear Science and Technology, Lanzhou University, Lanzhou 730000, China}
\author{Y. F. Niu}
\email{niuyf@lzu.edu.cn}
\affiliation{School of Nuclear Science and Technology, Lanzhou University, Lanzhou 730000, China}
\author{G. Col\`{o}}
\affiliation{Dipartimento di Fisica, Unversit\`{a} degli Studi di Milano, via Celoria 16, I-20133 Milano, Italy}
\affiliation{INFN, Sezione di Milano, via Celoria 16, I-20133 Milano, Italy}

\date{\today}

\begin{abstract}

The direct $\gamma$-decays of
the giant dipole resonance (GDR) and the giant quadrupole resonance
(GQR) of $^{208}$Pb to low-lying states are investigated
by means of a microscopic self-consistent model. The model considers effects beyond the linear response approximation. The strong sensitivity of $\gamma$-decay to the isospin of the involved states is proven. By comparing their decay widths,
a much larger weight of the
$3_{1}^{-}$ component in the GQR wave function of $^{208}$Pb is deduced,
with respect to the
weight of the $2_{1}^{+}$ component in the GDR wave function.
Thus, we have shown that
$\gamma$-decay is a unique probe of the resonance wave functions, and a
testground for nuclear structure models.

\end{abstract}

\maketitle

Collective excitation modes of many-body systems exist in many branches of physics.
In atomic nuclei the giant resonances (GRs) show up around 10-30 MeV \cite{Bortignon1998,Speth1981,Harakeh2001}, and they are
characterized by the
quantum numbers related to orbital angular momentum, spin, parity, and isospin.
Since the discovery of the first GR mode, i.e., the giant dipole resonance (GDR) in 1937 \cite{Bothe1937},
more and more GR modes have been discovered and studied.
Giant resonances are not only interesting in themselves, but one should also remind that
their properties can be linked to basic parameters of the nuclear equation of state (EoS)
\cite{ROCAMAZA201896}.
However, most of our well-established knowledge is still at the level of global properties,
such as  the mean resonance energy and the exhaustion of the appropriate
energy-weighted sum rule (EWSR). Despite past attempts, a more detailed understanding of the wave functions of the GRs, as well as of their decay properties, is still missing to a large extent.

Giant resonances are characterized by a large decay width $\Gamma$ of several MeV.
$\Gamma$ consists of several components.
The first one is the Landau width $\Delta \Gamma$, resulting from
the fragmentation of elementary one-particle-one hole ($1$p-$1$h) excitations,  that do
not all couple with themselves to form a single macroscopic collective state.
The second one is the escape width $\Gamma^{\uparrow}$, related to
the direct emission of nucleons, $\alpha$ particles, or photons.
The last one is the spreading width $\Gamma^{\downarrow}$,
 that constitutes as a rule
the dominant contribution to the total width
and arises from the coupling to $2$p-$2$h, \ldots, $n$p-$n$h states \cite{Bertsch1983,Drozdz1990}.

Although the above picture is widely accepted,
 the direct experimental evidences are not many. While the total width can be determined in
inclusive experiments, there are only few decay experiments that have tried to pin down the
precise values of $\Gamma^\uparrow$, and even less able to extract
the branching ratios $\Gamma^\uparrow_c/\Gamma^\uparrow$
associated to specific decay channels $c$ (cf. Table 8.3 of \cite{Harakeh2001}, as well as \cite{Hunyadi} and
references therein). There
have been several high-resolution experiments, but microscopic interpretations have been scarce.
One exception is the series of experiments carried out by the Darmstadt group that have been complemented
by a wavelet analysis of the underlying energy scales. The comparisions with the scales that emerged
from theoretical models has allowed to draw a few tentative conclusions.
It has been shown that, for the giant quadrupole resonance (GQR) in $^{208}$Pb,
the coupling of $1$p-$1$h configurations to the low-lying phonons is responsible for the fine
structure \cite{Shevchenko2004}, while for the GDR in $^{208}$Pb a different mechanism, i.e.,
Landau damping at the $1$p-$1$h level, leads to the same fine structure \cite{Poltoratska2014}.

$\Gamma_{\gamma}$ is a tiny part of the total decay width,
but since only the well-known electromagnetic interaction is involved,
the results can be more clearly interpreted than all those in which
the strong interaction is involved \cite{Bracco2019}.
The direct decay branch ought to be clearly distinguished from the decay through the compound
nucleus \cite{BEENE198519}.
Then, the $\gamma$-decay is extremely sensitive to the multipolarities of GRs \cite{Beene1989,Beene1990},
 and its study
provides a useful way to shed light on the microscopic properties of the GRs.
 Although these ideas have been put forward in the past,
the quantitative studies are also very rare.
Some of them were made when nuclear structure models were purely phenomenological \cite{Bortignon1984}, so that it
was hard to assess if $\gamma$-decay was a strong benchmark to test theories. Instead, here, we use
Skyrme functionals self-consistently.
Thus, the main goal of our
work is to show quantitatively how the $\gamma$-decay can give access to
microscopic properties of GRs and, in this respect, how it can be a testground for nuclear models.

The experimental data on the $\gamma$-decay of GRs to low-lying states are very rare due to the the small
branching ratio.
A few data were reported several decades ago \cite{Beene1989,Beene1990}; recently, new data have been
reported in Ref. \cite{Isaak2013}. However,
with the development of new $\gamma$-beam facilities like ELI-NP in Europe \cite{Tanaka2020}
and SLEGS in China \cite{Guo2008}, as well as with the advancing of $\gamma$-ray detectors
like ELIGANT-GN \cite{Krzysiek2019}, the measurement of $\gamma$-decay to low-lying states becomes more and
more feasible, so that new experiments are planned at ELI-NP \cite{Krzysiek2019} and RCNP \cite{proposal_E498}.

Under such circumstances, the theoretical study on the GR decay to low-lying states
 needs a revival. Existing theoretical works are either very old or, as we mentioned, purely based on
phenomenological input.
The surface coupling model \cite{Bortignon1984},
the extended theory of finite Fermi systems (ETFFS) \cite{Speth1985,Kamerdzhiev2004},
and the quasiparticle-phonon model (QPM) \cite{Ponomarev1992},
have been employed so far.
All these models incorporate effects
beyond
the mean-field level.

Only a few years ago, the first fully self-consistent treatment of
the $\gamma$-decay of GRs, based on the random phase approximation plus
particle-vibration coupling (RPA+PVC) model,
has become available
\cite{Brenna2012}.
It has to be noted that
the RPA+PVC model has been successfully applied
to study different nuclear properties, ranging from
the single-particle nuclear levels \cite{Litvinova2006,Colo2010,Litvinova2011,Colo2017},
the Gamow-Teller response and the related $\beta$-decay \cite{Niu2014,Niu2015,Niu2018,Robin2019,Litvinova2020},
to the spreading widths of non charge-exchange GRs \cite{RocaMaza2017,Shen2020}.
Therefore, in this work, we use the RPA+PVC model to investigate the $\gamma$-decay of GRs
to low-lying states
through electric dipole ($E1$) transitions
 and provide
useful guidance for possible experiments in the future.
The novelty of our work, with respect to previous ones, consists in having been able to
interpret the theoretical
results for the $\gamma$-decay in terms of basic properties of the
involved states (in particular, their isospin), and in terms of their microscopic wave function.

\begin{figure}[htbp]
  \centering
  \includegraphics[width=0.48\textwidth]{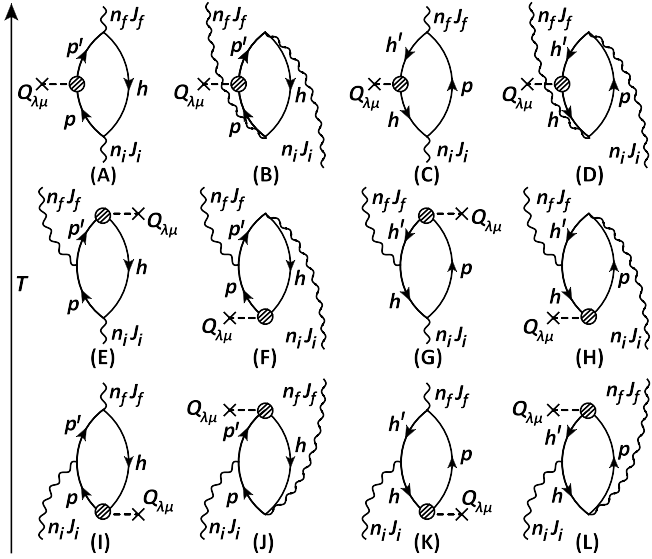}\\
  \caption{The 12 lowest order NFT diagrams in the process of $\gamma$-decay between two vibrational states.
           The circle with lines includes the contribution to $Q_{\lambda\mu}$ from
           nuclear polarization \cite{Brenna2012}.
           The arrow of time is upward.}
  \label{Fig1}
\end{figure}

The $\gamma$-decay width $\Gamma_{\gamma}$ is calculated as \cite{Bohr1998I}
\begin{equation}
   \Gamma_{\gamma}(E\lambda;i\rightarrow f)
    = \frac{8\pi(\lambda+1)}{\lambda\left[(2\lambda+1)!!\right]}
      \left(\frac{E}{\hbar c}\right)^{2\lambda+1}
      B(E\lambda;i\rightarrow f),
   \label{Eq_width}
\end{equation}
where $E$ represents the transition energy, and $\lambda$ is the transition multipolarity. The transition probability $B$ is
\begin{equation}
  B(E\lambda;i\rightarrow f)
  =\frac{1}{2J_i +1} \left| \left< J_f|| Q_{\lambda} ||J_i \right> \right|^{2},
  \label{Eq_BEL}
\end{equation}
where the initial and final vibrational states $|nJ\rangle$, denoted by a wavy line in Fig.~\ref{Fig1},
are calculated with the fully self-consistent RPA method \cite{Colo2013}.
In general, the electric multipole operator reads
\begin{equation}
\begin{aligned}
      Q_{\lambda\mu}
    =& \frac{e}{2} \sum_{i=1}^{A}
    \Big\{
    \Big[
    \Big(
    1-\frac{1}{A}
    \Big)^{\lambda}
    +(-)^{\lambda} \frac{2Z-1}{A^{\lambda}}
    \Big] \\
    &-
    \Big[
    \Big(
    1-\frac{1}{A}
    \Big)^{\lambda}
    + \frac{(-)^{\lambda+1}}{A^{\lambda}}
    \Big] \tau_{z}(i)
    \Big\} r_{i}^{\lambda} i^{\lambda} Y_{\lambda\mu}(\hat{\bm{r}}_i) \\
    \equiv& \frac{1}{2}\sum_{i=1}^{A}
     e^{\textrm{eff}}_{i}
     r_{i}^{\lambda} i^{\lambda} Y_{\lambda\mu}(\hat{\bm{r}}_i),
\end{aligned}
\label{Eq_Q_lm}
\end{equation}
where the effective charge caused by the recoil of the nucleus
has been introduced \cite{Bohr1998I}.

For the decay between two vibrational states, the RPA method, which describes  well only the transitions
between states that differ only by one vibrational phonon, is not enough.
 We have to consider the interplay between
collective phonons and individual particles,
i.e., the
particle-vibration coupling (PVC) effects. This
can be dealt with in a perturbative approach, by including all the lowest-order perturbative diagrams involving
single-particles states and phonons
like in the nuclear field theory (NFT) \cite{Bortignon1977,Bohr1998II}.
The 12 lowest-order NFT diagrams, as sketched in Fig.~\ref{Fig1}, are used in the calculations of
the reduced matrix element $\left< J_f|| Q_{\lambda} ||J_i \right>$.
Diagrams A-D are at the RPA level, while diagrams E-L are at the PVC level
due to the PVC vertex, e.g., $\langle p',nJ|V_{\textrm{res.}}|p\rangle$ in diagram E.
Diagrams E, F, G, and H will contribute when the initial phonon has more complex configurations made up with $1$p-$1$h
coupled with the final phonon, while diagrams I, J, K, and L will contribute
when the final phonon has more complex configurations made up with $1$p-$1$h coupled with the initial phonon.
For the detailed expressions, we refer to Ref. \cite{Brenna2012}.

\begin{table*}[htbp]
\centering
\caption{
Energies $E$ and electric transition probabilities $B(E\lambda)$ of the first $2^{+}$ and $3^{-}$ states,
as well as centroid energies and electric transition probabilities $B(E\lambda)$ of the GDR and GQR in $^{208}$Pb.
In the last two columns,
the $\gamma$-decay widths $\Gamma_{\gamma}$ of the GDR to the $2^+_1$ state, as well as of the GQR to the $3_1^-$ state,
are listed respectively. The experimental values are
from Refs.~\cite{Martin2007,Beene1989,Veyssiere1970}.}
\begin{tabular}{ccccccccccc}
  \hline
  \hline
   $^{208}$Pb & ~~~~$E_{2_1^{+}}$~~ ~~& ~~$E_{3_1^{-}}$ ~~& ~~$E_{\rm GDR}$ ~~&~~ $E_{\rm GQR}$~~
 & $B(E2\uparrow)_{2_1^{+}}$  & $B(E3\uparrow)_{3_1^{-}}$ & $B(E1\uparrow)_{\rm GDR}$ & $B(E2\uparrow)_{\rm GQR}$
 & $\Gamma_{\gamma}({\rm GDR}\rightarrow 2_1^+)$ & $\Gamma_{\gamma}({\rm GQR}\rightarrow 3_1^-)$ \\ \hline
   Units & MeV & MeV & MeV & MeV  & $10^{3}e^{2}{\rm fm}^{4}$ & $10^{5}e^{2}{\rm fm}^{6}$ & $e^{2}{\rm fm}^{2}$
         & $10^{3}e^{2}{\rm fm}^{4}$ & eV & eV\\ \hline
   Exp.       & 4.09        &   2.62      &~~13.5$\pm$0.1~~&~~10.9$\pm$0.3~~& ~~3.18$\pm$0.16~~ & ~~6.11$\pm$0.09 ~~
   &  $62.5\pm 5.0$ & ~~5.80$\pm$1.60~~  & ----- &   5$\pm$5 \\ \hline
   LNS\cite{Cao2006}          & 4.79        &   2.91      &  13.91        & 12.03       & 3.08          & 6.52          &  66.5  &  4.46
    &285.74 & 123.54\\ \hline
   SAMi \cite{RocaMaza2012}  & 4.03        &   2.88      &  13.36        & 12.20       & 1.31          & 7.65          &  72.9  &  5.51
    &184.43 & 112.95\\ \hline
   SkM$^{\ast}$\cite{Bartel1982}  & 4.86        &   3.22      &  13.65        & 11.63       & 2.91          & 5.77          &  73.1  &  4.89
    &307.48 &  79.40\\ \hline
  \hline
\end{tabular}
\label{Tab_1}
\end{table*}

The present calculation is done in a configuration space where the cutoff energy for single-particle
levels is $E_{\rm{cut}} = 150$ MeV, and the box size for  calculating the single-particle levels is 20 fm.
The EWSRs satisfy the double commutator values at the level of more than $99.50\%$.
The decay width is sensitive to the properties of the initial state $|n_i J_{i}\rangle$
and final state $|n_f J_{f}\rangle$.
Therefore, as a first step, we check the quality of the description of the low-lying states $2_1^+$ and $3_1^-$,
as well as that of the GDR and GQR,  by comparing energies and electric transition probabilities
with the experimental values.  These latter are listed in Table~\ref{Tab_1} together with the theoretical
results obtained using three of the many Skyrme sets, namely
LNS \cite{Cao2006}, SAMi \cite{RocaMaza2012} and SkM* \cite{Bartel1982}.
Among these 3 interactions, LNS describes well both the energies and electric transition
probabilities for low-lying states and GRs, since the largest discrepancy with experimental
data is less than $20\%$.
In particular, for the $E_{\textrm{GDR}}$ and $B(E\lambda)$ of low-lying states,
the discrepancies are not larger than $5\%$.
The calculated $B(E1\uparrow)_{\rm GDR}$ and $B(E2\uparrow)_{\rm GQR}$ also lie within the experimental error.
Therefore, it is reliable to use LNS in the following investigation of the $\gamma$-decay
from GRs to low-lying states. However, by checking the results of other Skyrme interactions,
we found that the main conclusion of this paper is independent of the choice of the interaction.

\begin{figure}[htbp]
\centering
\includegraphics[width=0.4\textwidth]{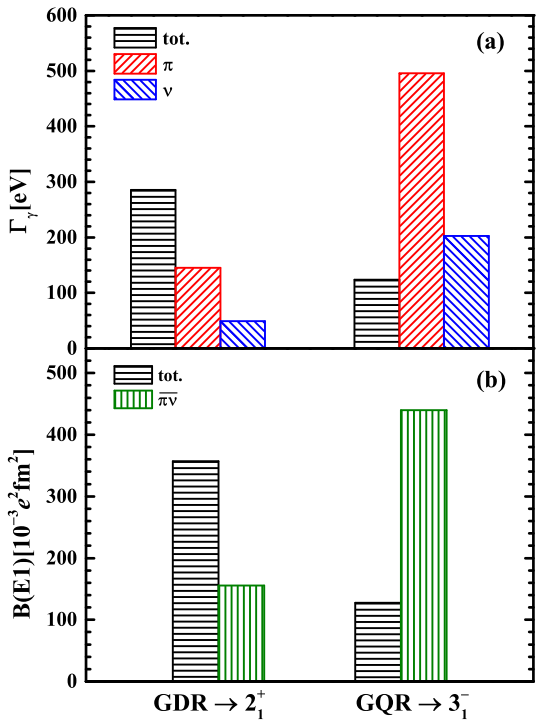}\\
\caption{Comparison of the $\gamma$-decay widths between GDR $\rightarrow 2_1^+$ and GQR $\rightarrow 3_1^-$
in $^{208}$Pb [panel (a)].
The total $\gamma$-decay widths (tot.) are shown, as well as the contributions from protons ($\pi$) and neutrons ($\nu$).
Comparison of the transition probabilities $B(E1)$ between GDR $\rightarrow 2_1^+$ and GQR $\rightarrow 3_1^-$
in $^{208}$Pb [panel (b)]. Here, the total $B(E1)$ and the average of
contributions from protons and neutrons $B^{\overline{\pi\nu}}(E1)$ are displayed.}
\label{Fig2}
\end{figure}

The calculated $\gamma$-decay widths $\Gamma_{\gamma}$ from GRs to low-lying states are
listed in the last two columns of Table \ref{Tab_1},
where the width of GDR $\rightarrow 2_1^+$ is 285.74 eV,
more than twice  than that of GQR $\rightarrow 3_1^-$, which is 123.54 eV.
Notice that the averaged transition energies for these two cases are both 9.12 MeV, which does not cause much difference in the decay widths.
To understand this difference, we first notice that the isospin properties of GDR and GQR are quite different,
 as the GDR (GQR) is mainly an isovector (isoscalar) resonance.
Therefore, we consider separately
the contributions from protons ($\pi$) and neutrons ($\nu$) to the total $\gamma$-decay widths,
that are shown in panel (a) of Fig. \ref{Fig2}.
One can see that for the GDR, the total decay width $\Gamma _{\gamma}$ is larger than the decay widths
stemming from only protons ($\Gamma_{\gamma}^{\pi}$) or neutrons ($\Gamma_{\gamma}^{\nu}$), that is,
$\Gamma _{\gamma} > \Gamma^{\pi(\nu)}_{\gamma}$. For the GQR, on the contrary, $\Gamma _{\gamma}
< \Gamma^{\pi(\nu)}_{\gamma}$. This means that protons and neutrons contribute coherently
to enhance the $\gamma$-decay width in the transition GDR $\rightarrow 2_1^+$, while they cancel
each other to reduce the $\gamma$-decay width in the transition GQR $\rightarrow 3_1^-$. This behavior can be explained by
 considering that the low-lying states $2_1^{+}$ and $3_1^{-}$ have both isoscalar character and that the $E1$
operator is isovector.
As a result, under the action of the $E1$ operator, the proton and neutron contributions have the same
phase in the GDR $\rightarrow 2_1^+$ case , but opposite phase for the GQR $\rightarrow 3_1^-$ case, so that
these decays are, respectively, either enhanced or suppressed \cite{Bortignon1984,Speth1977}.

\begin{table}[htbp]
\centering
\caption{Total $\gamma$-decay widths $\Gamma_{\gamma}$
         and contributions from protons ($\Gamma_{\gamma}^{\pi}$) and neutrons ($\Gamma^{\nu}_{\gamma}$)
         for GDR $\rightarrow 2_1^+$ and GQR $\rightarrow 3_1^-$, respectively, in $^{56}$Ni
         without the Coulomb interaction. Units are in eV.}
\begin{tabular}{cccc}
  \hline\hline
  $^{56}$Ni& $\Gamma_{\gamma}$ & $\Gamma_{\gamma}^{\pi}$  & $\Gamma^{\nu}_{\gamma}$ \\ \hline
  ${\rm GDR \rightarrow 2_{1}^+}$      &3877.72 & 969.43 & 969.43 \\
  ${\rm GQR \rightarrow 3_{1}^-}$      &   0.00 &  84.95 &  84.95 \\
  \hline\hline
\end{tabular}
\label{Tab_2}
\end{table}

We can further prove the above argument in the
limit of exact isospin symmetry. We pick up the $N=Z$ nucleus $^{56}$Ni and turn off the Coulomb interaction.
In this case, the $2_1^+$ and $3_1^-$ states are purely isoscalar, while the GDR and GQR states are purely isovector
and isoscalar, respectively. The contributions to the $\gamma$-decay widths from protons and neutrons,
$\Gamma_{\gamma}^{\pi}$ and $\Gamma^{\nu}_{\gamma}$,
 are equal.
In this limit,
a neat relation is found, namely
\begin{eqnarray}
      \Gamma (\textrm{GDR}\rightarrow 2_{1}^{+})
   & =& 4 \cdot \Gamma^{\pi(\nu)}(\textrm{GDR}\rightarrow 2_{1}^{+}),  \\
      \Gamma (\textrm{GQR}\rightarrow 3_{1}^{-})
    & =& \Gamma^{\pi}(\textrm{GQR}\rightarrow 3_{1}^{-})  - \Gamma^{\nu}(\textrm{GQR}\rightarrow 3_{1}^{-}) \nonumber\\
    &=&  0.
\end{eqnarray}
In Table \ref{Tab_2}, we  show the results of our microscopic calculations that fully confirm the above
expectation, of
the largest possible coherence between neutrons and protons in the GDR decay and of their complete cancellation
in the GQR decay. This serves as a limiting example to which the situation in $^{208}$Pb may be compared.

However, once we exclude the influence of the different isospin properties of GRs, and only compare
$\Gamma_{\gamma}^{\pi}$ or $\Gamma_{\gamma}^{\nu}$ in the two cases, GDR $\rightarrow 2_1^+$ and
GQR $\rightarrow 3_1^-$ in $^{208}$Pb, a completely different  pattern is observed.
Specifically, $\Gamma^{\pi}_{\gamma}$ and $\Gamma^{\nu}_{\gamma}$ in the decay of the GDR are, respectively, 145.11 eV
and 49.15 eV, while in the decay of the GQR they are both 3-4 times larger, namely 495.63 eV and 202.42 eV.

 In order to understand this relation, we firstly rule out the influence of the transition energies $E$
that enters $\Gamma_\gamma$,
and we directly use the total transition probability $B(E1)$ as well as the values of $B^{\overline{\pi\nu}}(E1)$,
that is, the average of the contributions from protons and neutrons
[see panel (b) of Fig. \ref{Fig2}].
Similar relations as for the $\gamma$-decay widths are found for $B(E1)$: $B(E1)$ for the GDR decay
is about 2 times larger than for the GQR decay but, nevertheless, for $B^{\overline{\pi\nu}}(E1)$
the value for the GDR decay is less than $\frac{1}{2}$ of the value for the GQR decay:
\begin{subequations}
\begin{eqnarray}
   B (E1;\textrm{GDR}\rightarrow 2_{1}^{+})
    &>& 2 \cdot B (E1;\textrm{GQR}\rightarrow 3_{1}^{-}),
    \label{BE_GDR-GQR1} \\
   B^{\overline{\pi\nu}}(E1;\textrm{GDR}\rightarrow 2_{1}^{+})
    &<& \frac{1}{2} \cdot B^{\overline{\pi\nu}}(E1;\textrm{GQR}\rightarrow 3_{1}^{-}).
    \label{BE_GDR-GQR2}
\end{eqnarray}
\end{subequations}
The latter of these two relations must have to do with the
wave functions of the initial and final phonon.

As already mentioned, the $B(E1)$ from the GRs to low-lying states
obtains contributions from 12 different amplitudes. The associated diagrams are sensitive to different
components of
the wave function of the initial or final phonon.

\begin{figure}[htbp]
  \centering
  \includegraphics[width=0.4\textwidth]{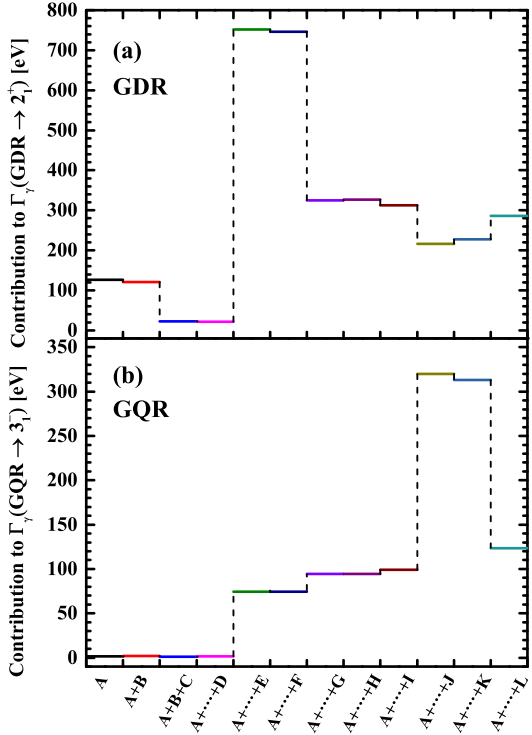}\\
  \caption{The contribution of each NFT diagram to the $\gamma$-decay widths
          $\Gamma_{\gamma}$ of GDR $\rightarrow 2_1^+$ [panel (a)] and GQR $\rightarrow 3_1^-$ [panel (b)] in $^{208}$Pb.}
  \label{Fig3}
\end{figure}

In Fig. \ref{Fig3}, we show the contribution of each of the 12 diagrams to the $\gamma$-decay widths
$\Gamma_{\gamma}$ of GDR $\rightarrow 2_1^+$ [panel (a)] and GQR $\rightarrow 3_1^-$ [panel (b)].
In the case of the GDR decay [panel (a)], diagram A, representing the particle contribution at the RPA level,
and diagram C, representing the hole contribution at the same level, almost cancel each other; hence,
only a small contribution remains at the RPA level (at which the wave function of both phonons is a
superposition simply of $1$p-$1$h configurations).
Once diagram E, representing  one of the particle contributions at the PVC level, is considered,
the width becomes very large;
diagram G, representing the corresponding hole contribution at the PVC level, cancels around $60\%$ of the width
that becomes 324.65 eV. Diagrams E, F, G and H represent the contributions (at PVC level)  that arise when
the wave function of the initial phonon has the component $|[(p h)_{J}\otimes J_f]_{J_i}\rangle$, namely
the coupling of $1$p-$1$h configurations with the final phonon.
The contributions from the diagrams I, J, K and L are small: these diagrams
represent the contributions (at the PVC level) that arise when the wave function of the
final phonon has the component $|[(p h)_{J} \otimes J_i] _{J_f}\rangle$, which corresponds to
the coupling of $1$p-$1$h configurations with the initial phonon.
 It is understandable that low-lying states can be coupled with $1$p-$1$h configurations
and be admixed with GRs that lie at similar energies. Having a high-lying GR plus a $1$p-$1$h states in the wave function of
a low-lying state is far more unlikely.
In the end, diagrams E and G yield $76\%$ of the total width.

For the GQR decay case in panel (b), all the diagrams at the RPA level are extremely small.
When diagram E is taken into account, $\Gamma_{\gamma}$ is 74.24 eV. When diagram G is added, it becomes 94.47 eV.
Diagram J contributes much to the width, but diagram L almost cancels it. Finally, we get 123.54 eV.
Similar to the GDR case, diagrams E and G yield $95\%$ of the total width.

As a conclusion, for both the GDR and GQR decay, diagrams E and G dominate the decay width.
It means that in this study of $\gamma$-decay, the component of $1$p-$1$h coupled with the final phonon
in the wave function of the initial phonon plays an essential role. Now,
from the discovery that $B^{\overline{\pi\nu}}(E1)$ of GQR $\rightarrow 3_1^-$ is larger than that of GDR $\rightarrow 2_1^+$,
as stated in Eq.~(\ref{BE_GDR-GQR2}), we can conclude that the $|[(p h)_{J} \otimes 3_{1}^{-}]_{\textrm{GQR}} \rangle$
component in the wave function of the GQR
is much larger than the $ |[(p h)_{J} \otimes 2_{1}^{+}]_{\textrm{GDR}} \rangle$ component in the wave
function of the GDR.
This conclusion is
in agreement
with that of the wavelet analysis, that we mentioned in the first part of this Letter.
Nevertheless, as we have already stressed, we have demonstrated that the analysis of
the $\gamma$-decay to low-lying vibrational states provides
a more clear and direct way to draw this conclusion.

In summary, the $\gamma$-decays from GRs to low-lying states in $^{208}$Pb are studied with the RPA+PVC model
by calculating the lowest order NFT diagrams.
First, we have proven the strong sensitivity of $\gamma$-decay to the isospin of the involved states.
Indeed, the decay GDR $\rightarrow 2_{1}^{+}$ is
isospin-enhanced while the decay GQR $\rightarrow3_{1}^{-}$ is isospin-suppressed.
If we exclude the isospin effects and consider the proton-neutron
average of the transition probabilities,
$B^{\overline{\pi\nu}}(E1)$,
this is found to be two times larger
in the case of GQR $\rightarrow 3_{1}^{-}$
than in the case of GDR $\rightarrow 2_{1}^{+}$.
This points clearly to a larger weight of the
$|[(p h)_{J} \otimes 3_{1}^{-}]_{\textrm{GQR}}\rangle$ component in the GQR wave function
with respect to the
$|[(p h)_{J} \otimes 2_{1}^{+}]_{\textrm{GDR}}\rangle$ component in the GDR wave function.

This work shows
{\em explicitly}
that the $\gamma$-decay of GRs to low-lying vibrational states is an effective approach
to access directly the microscopic structure of the GRs.
The same kind of study can be extended to other cases of interest like, for instance, that of the so-called
pygmy resonances (PRs) in order to understand to which extent they have a specific nature that makes
them different from GRs. In general, we look forward to new experimental data for
$\gamma$-decay, to answer more directly and systematically to questions like:
what are the microscopic structure and damping mechanism of GRs and PRs,
what are their isospin properties, how collective a PR is, and all the like.

\begin{acknowledgments}
This research is supported by the Fundamental Research Funds for the Central Universities under Grant No. Lzujbky-2019-11, and the European Union's Horizon 2020 research and innovation program under Grant No. 654002.
\end{acknowledgments}


\end{document}